# EVALUATION OF AN IMPULSE GRAVITY GENERATOR BASED BEAMED PROPULSION CONCEPT

**Chris Y. Taylor**[*]
Jupiter Research Corp.
Houston, TX
chrisytaylor@yahoo.com

**Giovanni Modanese**
University of Bolzano – Industrial Engineering
Via Sernesi 1
39100 Bolzano, Italy
giovanni.modanese@unibz.it

California Institute for Physics and Astrophysics
336 Cambridge Ave.
Palo Alto, CA 94306

## ABSTRACT

Recent experiments on anomalous forces involving high critical temperature superconductors by Podkletnov and Modanese have demonstrated an experimental apparatus apparently capable of generating a beam of radiation that acts on interposed objects as a repulsive force field. This beam is able to penetrate bulk material and imparts a force that is proportional to the mass of the target and independent of the target composition. Because of the similarity of these beam properties to gravitational effects and the possibility that the mechanism of its generation may be related to gravitation, this apparatus has been dubbed an "impulse gravity generator." Unlike gravitation, however, this beam has clear-cut boundaries and does not appear to diverge over the distance tested thus far (approximately 1 km). A beam of radiation or particles with the properties described for the impulse gravity generator would appear to be an excellent candidate for use in beamed spacecraft propulsion. Besides the usual benefits of beamed propulsion, it would not need sails or other special spacecraft components to function, could safely provide high accelerations to delicate components, and might operate at higher efficiencies than other beamed propulsion concepts. This paper analyzes the suitability of a beamed propulsion concept having properties consistent with the impulse gravity generator described by Podkletnov *et al*. The use of this propulsion concept for orbital maneuver, Earth-to-orbit, interplanetary, and interstellar applications based on presently available experimental results and theory is considered, and areas for future research needed to better characterize this phenomenon are discussed.

## INTRODUCTION

A conventional rocket must carry its engine, energy source, and reaction mass along with it. Beamed propulsion concepts seek to do away with one or more of these items on the spacecraft and replace them with a beam of radiation or matter, which is sent to the spacecraft from a transmitter that remains near the launch site. Eliminating these systems, their subsystems, and associated structure from the spacecraft reduces the amount of energy needed to propel the spacecraft to its destination. Because portions of the spacecraft's propulsion system are left at the launch site, they do not have to be built to the same exacting and expensive standards as the rest of the spacecraft. In addition, the transmitter components can be accessed relatively easily if repairs need to be made to them during flight; a feat that is often impossible with conventional rocket based spacecraft propulsion components. For missions with long acceleration times it may even be possible to upgrade the transmitter components during the flight. Also, because the transmitter remains behind after the initial mission is over it may be reused for future missions; becoming more like part of a space access infrastructure than just a spacecraft subsystem. Previously described beamed propulsion concepts include using a laser beam to heat rocket reaction mass,[1] pushing lightsail spacecraft with lasers[2] or masers,[3,4] and transferring momentum to spacecraft with a stream of high-speed pellets.[5] With all its advantages, beamed propulsion is likely to become an increasingly useful strategy for mankind's further exploration and exploitation of outer space.

Recent experiments on anomalous forces involving high critical temperature superconductors by Podkletnov and Modanese have demonstrated an experimental apparatus apparently capable of generating a beam of radiation that acts on interposed objects as a repulsive force field. This beam is able to penetrate bulk material and imparts a force proportional to the mass of the target and independent of the target composition. Because of the similarity of these beam properties to gravitational effects, and the possibility that the mechanism of its generation may be related to gravitation, this apparatus has been dubbed an "impulse gravity generator." This paper analyzes the suitability of a beamed propulsion concept using a transmitter that has properties consistent with the impulse gravity

---

generator described by Podkletnov *et al*. First, background information on Podkletnov's experiments and Modanese's theoretical explanation of the phenomenon will be provided. The general suitability of this phenomenon for beamed propulsion will be discussed with some conclusions on the likely characteristics of such a propulsion system. Following the description of general system characteristics, the concept is further illustrated by describing how it might be used in specific applications. Example applications analyzed in this paper include interstellar missions, interplanetary missions, Earth-to-orbit launch, and orbital maneuver of spacecraft near Earth. Finally, this paper highlights specific areas where further scientific research is needed to permit better analysis of impulse gravity generator based beamed propulsion, and to pave the way for its possible use in future spacecraft designs.

The technology behind the impulse gravity generator is not as mature and the theoretical model not as fully developed as most other beamed propulsion concepts, such as lasers, masers, and particle beams. Because of this lack of development, there are sometimes ambiguities in what the behavior of a mature impulse gravity generator beamed propulsion system would be. In situations where a lack of experimental results creates such ambiguities, educated assumptions will be made based upon the current theoretical explanations for the phenomenon developed by Modanese, frequently accompanied by a description of the effect on system behavior if the technology develops in a different way than predicted by current theory.

BACKGROUND

Impulse Gravity Generator Experiments
In their paper "Impulse Gravity Generator Based on Charged $Y Ba_2 Cu_3 O_{7-y}$ Superconductor with Composite Crystal structure", Dr. Evgeny Podkletnov and Dr. Giovanni Modanese describe a series of experiments involving an apparatus capable of exerting a brief repulsive force on small movable objects at a distance.[6] This apparatus is composed of a high-$T_c$ superconductor electrode from which a brief, high voltage electrical discharge is sent to a second electrode. This electrical discharge from the superconducting electrode is accompanied by the emission of a beam of anomalous forces that propagate in the same direction as the electrical discharge.

The anomalous beam generated by this apparatus has a diameter approximately the same as the superconducting emitter. The beam exhibits no detectible spread for distances of at least 1.2 km using measuring devices accurate to +/- 2 mm. Measurements beyond 1.2 km have not been made.

Dr. Podkletnov utilized target masses suspended in a pendulum arrangement within evacuated glass cylinders to measure the strength of the anomalous beam. The force exerted on the target masses increases with increasing discharge voltage. For any given voltage, the force on the targets is proportional to the mass of the target. Target composition and cross sectional area had no detectible effect on the force imparted by the beam. The beam propagates through different types of intervening material, including structural walls and electromagnetic shielding, without any noticeable attenuation in the force it exerts on the targets. The target masses used in the detector pendulums ranged from 10 to 50 grams. Target materials included metal, glass, wood, and rubber. Pendulum lengths of both 0.8 and 0.5 meters were used in the tests. At the maximum voltage reported, pendulum displacements of 5.6 inches (0.14 m) horizontally and 0.5 inches (.01 m) vertically where consistently measured for the target masses, representing on the order of $10^{-3}$ Joules of potential energy. Approximately $10^5$ Joules of electricity are consumed in one test run at this setting. Assuming the beam duration is approximately the same as that of the electrical discharge at the superconductor, such deflections would indicate a target acceleration of approximately 1000 gees (9810 m/s$^2$) for approximately $10^{-4}$ seconds. Since publication of these results, additional tests with target masses of up to 10 kg have been conducted with similar results.[7] Based on this information, the efficiency of the impulse gravity generator in these tests can be estimated to be at least $10^{-5}$ (0.001 %).

Many of the characteristics of the anomalous beam generated by this apparatus are suggestive of a gravitational nature, such as:
- Force exerted on the target is proportional to the target's mass
- Force exerted on the target is independent of the target's composition
- Beam penetrates bulk material
- Beam penetrates electromagnetic shielding

Based on its apparent gravitational nature, this anomalous beam has been dubbed a "gravity impulse" and will be referred to as such in this paper. Devices emitting this gravity impulse beam will be referred to as "impulse gravity generators."

Impulse Gravity Generator Theory
A brief analysis of the gravity impulse beam was published by Podkletnov and Modanese in 2001.[6] The starting point of this model is the concept of anomalous dipolar quantum fluctuations induced by the Cooper pairs condensate (see Appendix A). These fluctuations locally increase the gravitational vacuum polarizability to observable levels. In the weak gravitational shielding effect observed earlier by Pokletnov,[8,9] this increase was obtained in static or quasi-static form. In the impulse gravity generator, the critical conditions in the quantum condensate, necessary for the anomalous coupling between the gravitational field and the Cooper pairs, are produced only for a short time at the discharge. Therefore, it is natural to expect that in this case the formation of virtual dipoles be accompanied by the emission of virtual dipolar radiation with frequencies of the same order of the frequencies present in the discharge. A beam of virtual gravitons, whose momentum originates from that of the Cooper pairs of the discharge, is emitted in such a way that the beam direction is defined by momentum conservation. When the gravitons hit a target, they give rise to the observed impulsive forces.

It is hard to explain the observed features of this field and its propagation without resorting to the use of virtual radiation. The abnormal character of the observed radiation beam is readily apparent. It propagates through brick walls and metal plates without noticeable reduction in the beam's effect on the target masses. This is not, however, due to a weak coupling with matter, as is evidenced by the beam's effect on the targets. Furthermore, this beam conveys an impulse which is not related to usual momentum energy relationship for radiation, $E = cp$. Referring to the impulse gravity beam as "radiation" is therefore unsuitable, and one could possibly envisage it as an unknown quasi-static force field. In this way one could explain why an impulse is transmitted to the test masses and no shields are effective in stopping the effect of the beam.

The force appears to act upon several objects independently one from the other. For instance, if two pendulums are placed one behind the other along the beam, they move in the same way. This is just what would happen for a regular gravitational force produced by a source: it would act upon several objects without interference, the total reaction force on the source depending of course on how many and how massive "targets" are present. It is hard, however, to understand how such a field could be so well focused. There are no solutions of the classical field equations that allow for such a narrow and clear-cut force beam. Still, classically there exist wave-like solutions of the field equations. These solutions, however, cannot give such a beam either, and furthermore gravitational waves are quadrupolar and cannot exert any net force upon a target.

In quantum theory, any force is explained at a fundamental level as due to the exchange of virtual mediating particles. Technically, the expression for the interaction potential energy as the result of an exchange of virtual particles has been known for a long time in electrodynamics and in the gauge theories of weak and strong interactions. The corresponding formula for gravitation was given by Modanese,[10] where the static interaction potential of two gravitational sources is expressed as an integral of the graviton propagator over momenta and energies. It is also possible to prove that the main contribution to the integral comes from momenta and energies related by the equality $E<<cp$, which corresponds quite well to what has been observed in the experiment but differs from the energy-momentum relation for free massless particles which can propagate to infinite distance independently from the existence of a target. In this formula also appear, besides the integral over momenta and energies of the exchanged particles, factors corresponding to the two masses. This means that the flow of exchanged virtual gravitons does not have an "absolute" intensity, but is instead proportional to the source mass and to the target mass at the same time! Such a picture clearly differs from what one imagines in the case of a real flux of particles emitted from a source and absorbed by a target.

In addition to the strangeness of the field configuration as already discussed, the force is repulsive. This is not explainable in the classical realm either. From the quantum point of view, what matters in a scattering between two particles mediated by a third virtual particle (the simplest example is the scattering electron/proton, mediated by a virtual photon) is the exchanged energy-momentum. This is so true, that the signs of the electric charges, in the example above, do not affect the differential cross section. An electron bounces against a proton exactly the same way a positron would. This describes an elementary process where one single virtual particle is exchanged in a very short time interval. The static force is the opposite limit; many virtual particles are exchanged, in a continuous way, and in the time limit from - infinity to + infinity. The phenomenon underlying the impulse gravity generator represents a sort of intermediate case, because we are supposing that many gravitons are exchanged, but still with the features of a scattering or similar.

If the effect is gravitational, then the acceleration of a test mass should not depend on its mass or initial velocity, which causes an additional conceptual difficulty concerning power limitations. The power imparted to an accelerated spacecraft, or any other object, is given by the equation:

$$P_{spacecraft} = mav \qquad (1)$$

where *a* is the acceleration of the spacecraft, *m* is the mass of the object, *v* is the velocity of the spacecraft, and $P_{spacecraft}$ is the power added to the spacecraft in the form of increasing kinetic energy. If the mass or velocity of the target is too high, then the amount of power needed to maintain the expected acceleration will exceed the power available in the beam. If insufficient power is available in the beam to produce the expected acceleration for the target mass, then either the effect violates the equivalence principle, or more likely the concept of field breaks down when energy required exceeds that available (in the sense that there will be a back-reaction on the source). It may be possible that an impulse gravity beam operating in such a power-limited fashion might provide the expected amount of acceleration, but do so only to a portion of the mass in the beam path. The occurrence of such a condition in a spacecraft undergoing high acceleration in an impulse gravity beamed propulsion system could lead to very large internal stresses in the unaccelerated portion of the spacecraft and possible damage to or loss of the spacecraft.

If the impulse gravity beam is a phenomenon based on virtual particles flowing between the emitter and massive objects in the beam path, then the percentage of the beam's available energy imparted to a target may be dependent only on the mass of the target, with no dependence on the percentage of the beam path the target occupies. In such a case, a target that fills only a small percentage of the beam's cross sectional area might still convert most, or all, of the beam's available power into acceleration. This is clearly different than what occurs with a beam composed of real radiation or particles, which fill the beam and can be described as having a power per of beam cross sectional area. So far impulse gravity experiments have not shown the force imparted on the target to have any dependence on the percentage of the beam the target occupies or the amount of target mass per unit of beam cross sectional area. These results could be interpreted to mean that there is no relationship between the force on the target and target cross section, or that the experimental trials were all conducted at or below the amount of energy available in the beam section occupied by the target mass. Either conclusion will have significant, though different, implications for impulse gravity beamed propulsion practicality. If the former is true, then impulse gravity beam transmitters with long effective ranges could more easily be constructed, because large beam diameters resulting from beam divergence or deliberately created to compensate for pointing errors will not require a corresponding increase in beam power to compensate for the portion of the beam that "misses" the target. If the latter is the case, then it would suggest that the experimental trials thus far have still not approached the power per unit cross sectional area available in the current impulse gravity generator designs, and the current impulse gravity generator efficiencies are much higher than the low-end estimate made from existing data.

There is also the problem of how to explain the coherence in the beam. It resembles the coherence of a laser beam, even though in this case the beam is probably only virtual in the sense that its existence is limited in time and it does not have an absolute intensity. A strict analogy with lasers is impossible because the resonating cavity is missing. Lasers and masers, however, are only a particular example of systems that exhibit so-called self-organized behavior. These are systems made of a large numbers of particles, which evolve in time not strictly individually, but with coordinated dynamics. This in turn is possible due to a form of interaction-communication between the particles, which allows them to synchronize their actions. The general conditions necessary for the self-organized behavior to occur, are believed to be the following[11]:

1. a very large number of particles/atoms/emitters
2. an indirect interaction between them (in lasers this is the stimulated emission, favored by the presence of mirrors of the resonant cavity)
3. an energetic supply from the outside
4. a suitable geometric arrangement of the systems, which "suggests" a definite way of organization
5. a restless activity of the system in the sense of exploring all its possible configurations through random fluctuations

The impulse gravity generator satisfies conditions 1, 3, 4 and condition 5 may be satisfied by the possible presence of enhanced gravitational fluctuations. It is not yet clear what portion of the impulse gravity generator might satisfy condition 2.

If the impulse gravity beam propagates in air, some energy should be depleted from it as it propagates. At standard temperature and pressure, the energy lost to air from the beam generated by Dr. Podkletnov's experimental impulse gravity generator should be on the order of $10^{-3}$ J/m. The velocity shear between the air in the path of the impulse and surrounding atmosphere should, in principal, lead to noticeable air turbulence after the pulse. While extensive studies of the behavior of the air in the impulse gravity beam have not been conducted, observations of the air in the beam path with smoke show that only brief forward and back movement of the particles occurs. There is no significant airflow, as the impulse is very short, and there is no turbulence and no vortex phenomena. From the theoretical point of view, the action of the beam on air is not easy to describe. One can regard air as an elastic medium with pressure etc., or one can consider scattering of radiation by the single molecules, and it is impossible at this stage to tell which model is best. If we assume, as done above, that the impulse is not a classical gravitational field, but a beam of virtual gravitons with E<<pc, then it is possible to predict a small interaction with air. When one of these gravitons hits a gas molecule, it cannot be absorbed. In fact, the total energy and momentum must be conserved in the collision, but this is impossible because the relation $E=p^2/2m_{molecule}$ would have to hold after absorption. On the contrary, if a graviton hits a molecule of a solid the energy can be immediately redistributed to the whole body, so that after absorption $E=p^2/2m_{solid}$, which is much smaller. In technical terms, within a solid graviton absorption occurs "in an external field", while in a gas it is an elementary process forbidden by energy-momentum conservation. All this implies that the interaction of the gravity impulse may be dependent on the state of the matter in the beam path.

Impulse Gravity Beam Propulsion
The impulse gravity beam apparently has many characteristics that make it very attractive for use in beamed propulsion. One of the most obvious is that its effects are independent of target composition. Typical beamed propulsion concepts require some portion of the spacecraft be specially designed to capture the energy or momentum of the beam and convert it into propulsion. Clever design can reduce the mass penalty for such components by having them serve dual functions, like using a lightsail as part of a sensor or communication system, but not eliminate it completely. The impulse gravity beam can be used to accelerate a spacecraft without using any propulsion specific components onboard the spacecraft; the beam interacts directly with the mass comprising the spacecraft payload, structure, and systems. The elimination of any beamed propulsion components on the spacecraft can provide a substantial mass savings and a higher payload mass fraction for spacecraft designed with impulse gravity beamed propulsion in mind. Because impulse gravity beamed propulsion does not require any system specific components on the spacecraft, it can be easily incorporated into existing spacecraft designs and even used with spacecraft that have already been launched. Impulse gravity beamed propulsion could also be applied to completely inert and/or non-cooperative targets, such as asteroids or debris.

Beamed propulsion systems that require spacecraft components to assist in propulsion have acceleration limits imposed by those components, such as the sail thermal limit for laser pushed lightsails. By eliminating the need for such spacecraft components, the impulse gravity beamed propulsion concept also eliminates any limitations they may impose on acceleration. Because the impulse gravity beam penetrates bulk material and seems to act independently of target composition, it will uniformly accelerate all spacecraft components in the beam path. Even at high accelerations, the spacecraft components would not experience any internal stresses; a spacecraft being propelled by an impulse gravity beam would behave as though it were in freefall. Uniform acceleration of all spacecraft components means that even delicate payloads might safely undergo very high accelerations. Elimination of internal stresses caused by high acceleration of the spacecraft would also allow considerable structural mass savings. A spacecraft propelled by an impulse gravity beam might be little more than a collection of parts flying in close formation. The previously mentioned evidence regarding beam absorption in air and the possibility that the beam's effects might be state dependent suggests potential limitations to this benefit.

Accelerations of test masses at approximately 1000 gees of acceleration have been reported so far[6] and it is reasonable to assume that higher accelerations will be possible in the future. The benefit of high acceleration, beyond the obvious reduction in trip time, is to reduce the amount of time needed to achieve cruise speed. A shorter acceleration time means that there will be less problems arising from the need to compensate for the motion of the transmitter. For transmitters in planetary orbits, short acceleration times relative to the orbital period opens up more options in what orbits may be selected, since the transmitter no longer needs an unobscured line of sight to the spacecraft for the complete orbit. Short acceleration times might even allow transmitters to be placed on rotating celestial bodies, such as the Earth or the Moon, which could ease problems of both constructing the transmitter and managing the reaction force on the transmitter caused by the beam.

High spacecraft acceleration also reduces the distance required for the spacecraft to accelerate to its cruise speed. The reduction in acceleration distance for the spacecraft also means a reduction in the distance over which the beam must be sent. This eases the problem common to beamed propulsion concepts of making sure the spacecraft stays within the beam over the entire acceleration distance.

It is fortunate that the impulse gravity generator is capable high accelerations, and therefore short acceleration distances, because it may have a short effective range when compared with other beamed propulsion concepts. Unlike beams of electromagnetic radiation, there is no known method of focusing or steering the impulse gravity beam once it is generated. Considering the theorized nature of the beam, such manipulation may be impossible. If this is the case, then beam quality, beam pointing accuracy, and beam jitter can only be controlled at the emitter. If the impulse gravity beam does have a short range relative to other beamed propulsion concepts then it may still be an effective propulsion method by using high power transmitters to accelerate the spacecraft to cruising speed in a short distance and/or a relay arrangement of multiple transmitters strung out along the spacecraft's acceleration path.

There are several possible impulse gravity beam developments that could allow the construction of impulse gravity generators with operational ranges of interplanetary, or even interstellar, distances. Even if the beam divergence, jitter, and pointing accuracy can only be controlled at the emitter, that does not pose a range problem as long as the emitter conditions can be controlled accurately enough. If, as discussed earlier, the momentum delivered to the spacecraft is not limited by the percentage of the beam cross sectional area it occupies then beam divergence may not be a significant problem and poor pointing accuracy can be compensated for by increasing the beam diameter without needing to worry about increasing the total beam power to compensate for the portion of the beam that "misses" the spacecraft. Even if increasing beam diameter does require increasing power consumption, using a large diameter beam may still be an acceptable solution if power costs are sufficiently low and/or overall system efficiency is sufficiently high that the total cost is still competitive with other propulsion concepts. For cases where the target spacecraft starts out near the transmitter, a low pointing accuracy may be acceptable if the beam jitter can be kept low enough that the spacecraft can maneuver itself to stay inside the beam path. Lastly, methods may be developed that allow impulse gravity beams to be focused and diverted with the same ease that laser, maser, and particle beams are controlled. Increasing the effective range of the impulse gravity beam makes it a more attractive propulsion concept because it allows spacecraft to be accelerated to cruise speeds with lower transmitter power levels and it allows the beam to be used in applications where the target's initial position is not in the vicinity of the transmitter.

The power required to accelerate a mass increases linearly with the velocity of that mass, as described in equation 1. It is obvious from this equation that an impulse gravity beam providing a constant power and acceleration, or a pulsed impulse gravity generator providing a constant average power and average acceleration, to a spacecraft of constant mass must operate at increasing efficiency as the spacecraft's velocity increases. At some point, however, the power required to maintain this constant acceleration of the spacecraft will exactly match the amount of power available to the impulse gravity beam. Below this transition point the excess power supplied to the transmitter is not transferred to the target spacecraft and does not provide any propulsion benefit. Above this transition point the force being applied by the beam must decrease as the velocity increases or else the power supplied to the target as increasing kinetic energy would exceed the power available to the beam.

The power consumed by an impulse gravity beamed propulsion system can be described by the equation:

$$P_{transmitter} = \eta_{system} P_{spacecraft}$$

(2)

where $P_{transmitter}$ is the power consumed by the transmitter, $P_{spacecraft}$ is the increase in spacecraft kinetic energy per unit time, and $\eta_{system}$ is the overall efficiency of the beamed propulsion system. Overall system efficiency can be defined by the relation:

$$\eta_{system} = \eta_t \eta_b$$

(3)

where $\eta_t$ is the efficiency with which power consumed by the transmitter can be converted into power in the impulse gravity beam, and $\eta_b$ is the efficiency with which power available to the impulse gravity beam can be converted to an increase in kinetic energy of the target at the operating acceleration and target conditions. The value of all of these efficiencies for current impulse gravity generator designs is unknown. The only experimentally available information on the efficiency of an impulse gravity beam generator is data collected on the effect of the beam on stationary targets suspended in a pendulum arrangement. The efficiency figure given above of "at least $10^{-5}$ (0.001 %)" is the highest system efficiency demonstrated experimentally. There is no evidence available for what contributions transmitter efficiency ($\eta_t$) and beam coupling efficiency ($\eta_b$) make to the overall efficiency.* As mentioned previously, the beam coupling efficiency increases as the power needed to accelerate the test mass increases up to some limit, which has yet to be reached experimentally. Because the limits of the current generator designs have not been reached with current test masses, the efficiency attainable by current designs may be much higher than this lower limit indicates. Considering the theorized nature of the impulse gravity beam, it seems likely that under ideal conditions up to 100% of the energy available to the beam could be converted into spacecraft kinetic energy. This is in considerable contrast to beamed propulsion concepts based on electromagnetic radiation, where even in ideal circumstances only a small fraction of the beam's power goes toward increasing the momentum of the target spacecraft. While no experimental tests of the impulse gravity generator have been conducted on targets moving at high speed, the theorized nature of the impulse gravity beam suggests that with the proper transmitter power and acceleration settings efficient beam coupling could be obtained at any spacecraft velocity. This is also significantly different from most other beamed propulsion concepts, which achieve their best efficiencies only for certain velocity ranges (laser/maser pushed sails are most efficient at high spacecraft velocities, and pellet stream momentum transfer is most efficient at low spacecraft velocities). The transmitter efficiency ($\eta_t$) of current impulse gravity generators is very inefficient; much of the energy consumed in the generator is lost as waste heat, light, and noise. It is likely that less than 1% of the energy consumed by the impulse gravity generator is available to the impulse gravity beam. Despite these low transmitter efficiencies, the high beam coupling efficiency allows even the current impulse gravity generator to achieve a system efficiency that is competitive with other beamed propulsion concepts.

---

* Describing system efficiency as a function of transmitter efficiency and beam coupling efficiency would seem to suggest that the beam has some "absolute" intensity which is a function of transmitter power and transmitter efficiency, however current theory (as previously explained) describes the beam as consisting of a flow of virtual gravitons whose magnitude is dependent on both the transmitter and the target simultaneously and is no greater than the target mass(es) can absorb. Even though it is convenient in equation 3 to divide the system inefficiency into transmitter related losses ($\eta_t$) and target related losses ($\eta_b$) for the purposes of describing the source of power losses, care must be taken when attempting to apply that division to the actual transmitter/target behavior, as could be done with a real flux of particles.

For the target masses and initial velocities reported so far, the impulse gravity beam achieves system efficiencies thousands of times higher than a theoretically perfect laser pushed lightsail system.* It is reasonable to expect that as the technology matures past the experimental stage the efficiency of the impulse gravity beam generation process will be improved. Considering that most of the systems inefficiency is in the beam generation process, such gains would also lead to corresponding improvements to overall system efficiency.

## APPLICATIONS

### General

While a spacecraft using impulse gravity beamed propulsion does not need any beamed propulsion specific components, there are certain characteristics likely to be common in spacecraft specifically designed to use this propulsion method. The exact nature of these design characteristics will depend on impulse gravity beam system properties that have not yet fully been characterized. One design characteristic that impulse gravity beamed propelled spacecraft will not have is a sail or other specialized beam collection device. Such spacecraft would, therefore, have much fewer design restrictions imposed on them by the propulsion system than are ordinarily required by more conventional beamed propulsion concepts.

If the impulse gravity beam has a definite energy per unit of beam cross sectional area and the maximum amount of power that can be applied to the target spacecraft is limited by the amount of the beam cross sectional area that is occupied by the spacecraft, then the spacecraft will likely be designed to present as even of a mass distribution across the spacecraft profile as possible, when viewed from the point-of-view of the transmitter. This is because the transmitter will likely need to generate a beam with sufficient power to penetrate and propel even the densest portion of the spacecraft, in order to minimize internal stresses in the spacecraft due to differential propulsion being applied to different components. With such beam characteristics, it is desirable to have the spacecraft occupy as much of the beam path as possible; this will likely result in a roughly cylindrical shaped spacecraft that presents a circular profile to the transmitter. The specific diameter of the profile that the spacecraft presents to the beam source will depend on the beam divergence, pointing accuracy, and/or beam jitter of the impulse gravity beam, as well as the mission design requirements of the spacecraft.

---

* Assuming a sail of negligible mass and perfect reflectivity, with the laser acting perpendicular to the sail and parallel to the direction of travel.

If the percentage of the beam power that can be applied to the target spacecraft is not dependent on the percent of the beam cross section that the spacecraft occupies then the mass distribution of the spacecraft components is irrelevant to its suitability for impulse gravity beamed propulsion. Such beam characteristics would likely also result in a very large beam diameter relative to the size of the spacecraft because there would be no need to reduce the amount of beam "missing" the spacecraft and having a larger beam size would ease beam-pointing requirements for the transmitter. In this case, the need to present a circular profile to the transmitter would also be reduced because the irregularities of the spacecraft's profile would be smaller relative to the size of the beam diameter and the likelihood of a protrusion on the spacecraft leaving the beam path would be greatly reduced. With such beam characteristics, impulse gravity beamed spacecraft layout would more closely resemble conventionally propelled spacecraft designs than it would resemble other beamed propulsion spacecraft designs.

Just as the exact details of impulse gravity beam propelled spacecraft cannot yet be determined with existing information, there are many unknowns in what the exact characteristics of a mature impulse gravity beamed propulsion transmitter design will be. Existing impulse gravity generator technology only generates the impulse gravity beam for a very short period of time, on the order of $10^{-4}$ seconds. For a practical propulsion system, the transmitter will need to greatly increase the amount of time it provides propulsion to the target spacecraft. This increase might be achieved with the development of an impulse gravity generator that is able to operate in a steady state condition. If such a generator cannot be built, then pulsing one or more generators at a high frequency could still achieve a high average acceleration of the target spacecraft, even though each individual pulse may be of short duration.

Current impulse gravity beam emitters are made from a disk of high-temperature superconductor approximately 100 mm in diameter, and generate a beam diameter of the same size. Larger diameter emitters are possible. The emitter size needed for an impulse gravity beamed propulsion transmitter would depend on the amount of divergence in the beam and the distance between the transmitter and the target spacecraft. It would be very beneficial if a method could be found to control the impulse gravity beam divergence so that the optimum beam diameter could be maintained regardless of spacecraft distance. Another way to vary beam diameter could be to vary the emitter diameter. One way to accomplish this might be to construct a large emitter from several smaller emitters in close proximity. Beam size (and even shape) could be altered by using different combinations of the small emitters elements. This design could have the additional benefit of allowing the beam strength to vary across the width of the beam by applying different voltages to the individual emitter elements.

Equation 1 suggests that in order for an impulse gravity generator to maintain constant acceleration on a target spacecraft with the least amount of wasted beam power, it would be necessary to continually increase the power available to the impulse gravity beam so that it was always at, or just above, the power level needed to keep accelerating the target spacecraft. It is likely that in a real beamed propulsion system, the transmitter would have a maximum power it could operate at as a result of power production, storage, or control limitations. To achieve cruise speed for the target spacecraft in as short an acceleration distance as possible with such a power limited transmitter, it would be necessary to operate at maximum transmitter power and maximum efficiency possible. Accomplishing this would require varying the spacecraft's acceleration through the acceleration process, starting with a high acceleration and reducing the acceleration provided as the target spacecraft's speed increased, in order to maintain the system at or just above the transition point.

Interstellar
Beamed propulsion is commonly proposed for interstellar missions because the very large spacecraft velocities needed to carry out the mission in a reasonable amount of time are very difficult to obtain with the rocket based propulsion systems commonly used for missions to nearer destinations.

Many beamed propulsion systems have the advantage that they can also be used to decelerate the spacecraft when it approaches the target system to allow rendezvous missions.[12] It seems unlikely that impulse gravity based beamed propulsion system, however, would be useful for performing such remote spacecraft decelerations. Most remote beamed propulsion decelerations rely on reflecting the beam back at the payload section of the spacecraft from a sacrificial reflector that is ejected from the rest of the spacecraft, but currently there is no known way to reflect an impulse gravity beam and given the nature of the phenomenon such manipulation may be impossible. An impulse gravity beam that imparted a momentum change toward the emitter instead of away from it might be used to pull on the spacecraft for deceleration, however there is no current evidence that an attractive impulse gravity beam can be constructed. Another possible way in which an impulse gravity beam could be used to decelerate an interstellar spacecraft would be to use some other force, such as interactions with local magnetic fields[13] to allow the spacecraft to circle around behind the target star system until it is moving toward the transmitter so that the spacecraft could be decelerated with an unreflected, repulsive impulse gravity beam. For any of these methods to work, the impulse gravity beam propulsion system would need to have an effective range long enough to reach interstellar distances. It is not yet known if such long range is achievable. If a rendezvous interstellar mission is desired, then a spacecraft launched with impulse gravity based beamed propulsion may need to rely on some other technology to provide spacecraft deceleration. Considering the potential difficulty that an interstellar rendezvous mission may pose for impulse gravity beamed propulsion technology, a flyby mission with no spacecraft deceleration was chosen as the sample interstellar mission to be analyzed by this paper.

Because of the high spacecraft velocities that must be achieved, and the correspondingly large amounts of energy that need to be expended to achieve those velocities, the reduction of spacecraft mass will likely be a very important factor in planning future interstellar missions. Determining the optimum spacecraft mass for any real future interstellar mission will require considerable engineering analysis to balance to the desire for low mission cost with the need to achieve results that could not be accomplished with less ambitious methods, such as SETI or highly advanced space based observatories. In order to achieve the best results with a minimum of spacecraft mass, future interstellar missions may rely heavily on nanotechnology[14] or utilization of in-situ resources[15]. Without the necessary information to perform such a detailed analysis, an educated guess of 1 metric ton was chosen for the spacecraft mass of the sample flyby interstellar mission. A payload mass of 1 metric ton is a common assumption in other interstellar propulsion papers, so it should allow easy comparison between this and other papers describing alternative propulsion methods for interstellar missions

Like spacecraft mass, the determination of spacecraft cruising speed for an interstellar probe would likely be the result of complex, mission specific trade studies. For the purpose of this example, a cruising speed of 0.11c ($3.3 \times 10^7$ m/s) was chosen. This speed will just allow a flyby mission to Alpha Centauri, the nearest star system, 4.3 ly ($4 \times 10^{16}$ m) away to be accomplished within the fifty-year professional career that can be expected for humans.* Exact mission time will depend on the length of the acceleration phase, with a minimum average acceleration of 0.0076 gees (0.075 m/s$^2$) required to achieve the assumed 50-year mission requirement. A cruise speed of 0.11c is commonly used as the assumed minimum acceptable for an interstellar mission, so it should allow easy comparison between this and other papers describing alternative propulsion methods for interstellar missions.

The kinetic energy contained in a 1 MT spacecraft traveling at a cruising velocity of 0.11c is $5.49 \times 10^{17}$ J. The power required to accelerate this spacecraft to its cruising velocity will depend on the efficiency of the propulsion system and the amount of time (and therefore distance) required for the acceleration. This relationship can be described by the equation:

$$P_{ave} = \eta_{system} E_{spacecraft} / t$$

(4)

where $P_{ave}$ is the average power consumption of the propulsion system, $\eta_{system}$ is the overall efficiency of the propulsion system, $E_{spacecraft}$ is the total kinetic energy increase of the spacecraft needed for it to attain its cruising speed, and t is the amount of time for which the spacecraft is being accelerated. Given the amount of resources and technological development needed to attempt an interstellar mission, it is reasonable to assume that one will not be attempted in the near future. By the time interstellar missions do become economically and politically feasible, it is likely that several orders of magnitude improvements can be made in impulse gravity generator efficiencies, ideally giving overall efficiencies of at least .01 (1%). The effective range of the impulse gravity beam transmitter will likely influence the choice of spacecraft acceleration time heavily. Acceleration time does not need to be limited by transmitter range, however, because multiple short-range transmitters could be used in a relay-like fashion to allow longer acceleration distances and therefore increase the acceleration time of the spacecraft.

The trade off between transmitter power and acceleration distance for an interstellar mission will likely be a complex problem whose solution is highly dependent on the characteristics of a mature impulse gravity beamed propulsion system that we do not currently have the ability to predict. Figure 1 shows the average system power needed to accelerate the previously mentioned example interstellar flyby spacecraft to its cruise speed of 0.11c for a wide variety of possible system efficiencies and acceleration distances. The y-axis on the left side of the graph gives the power consumption of the transmitter system per kilogram of spacecraft mass, while the y-axis on the right side of the graph shows total power consumption for the 1 MT example spacecraft. To give a sense of scale for the power figures given, the average global power consumption for 1997[16] and the power generated by a Saturn V first stage are also indicated.

---

* Changes in life expectancy or professional habits by the time an interstellar mission is feasible may make this career length assumption invalid, however it will still probably be desirable to avoid launching an interstellar probe until mission times of approximately 50 years can be achieved in order to reduce the chance of the first, slow probe being beaten to its destination by a later, faster probe.

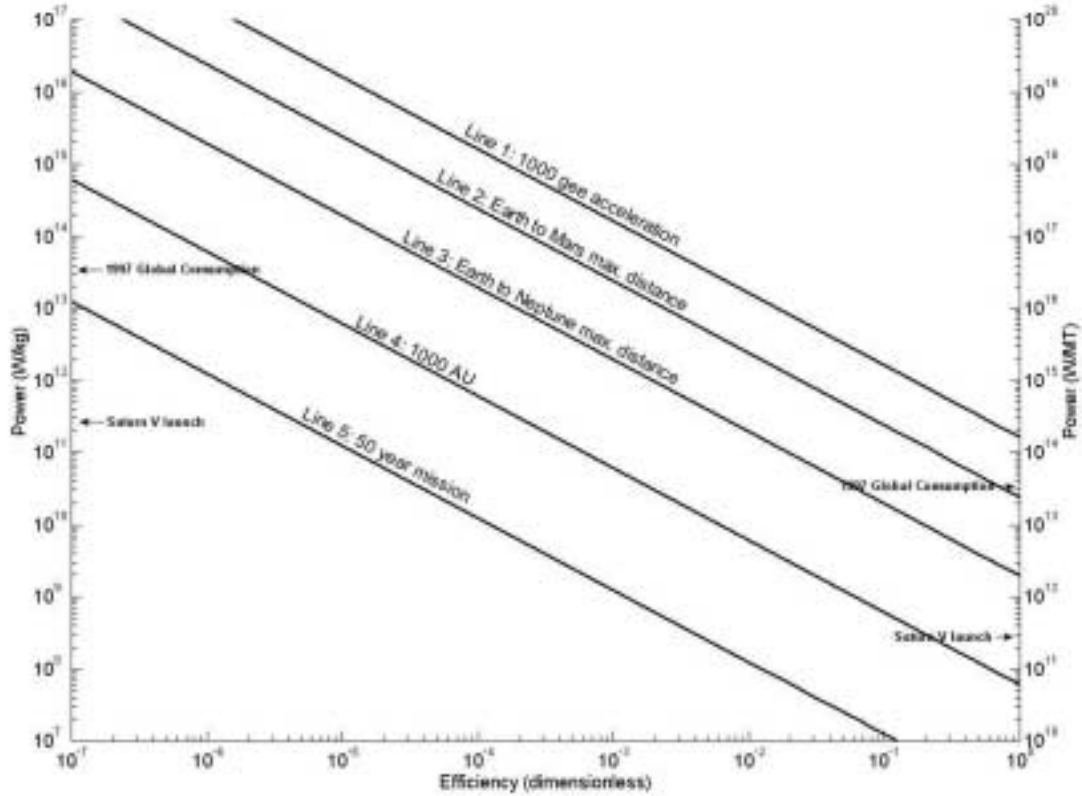

Figure 1: Average Power Consumption vs. System Efficiency for Impulse Gravity Beamed Propulsion of Spacecraft to 0.11c

| Line | Description | Acceleration Distance (m) | Ave. Acceleration (gees) | Acceleration Time (s) | Acceleration Time (days) |
|---|---|---|---|---|---|
| 1 | 1,000 gees | 5.6E+10 | 1000 | 3364 | 0.04 |
| 2 | Earth to Mars | 3.8E+11 | 147 | 22884 | 0.26 |
| 3 | Earth to Neptune | 4.7E+12 | 12 | 281,793 | 3.26 |
| 4 | 1000 AU | 1.5E+14 | .37 | 9.1E+6 | 105 |
| 5 | 50 year mission | 7.3E+15 | .0076 | 4.4E+8 | 5100 |

Table 1: Acceleration Time for Selected Distances for a 0.11c Spacecraft

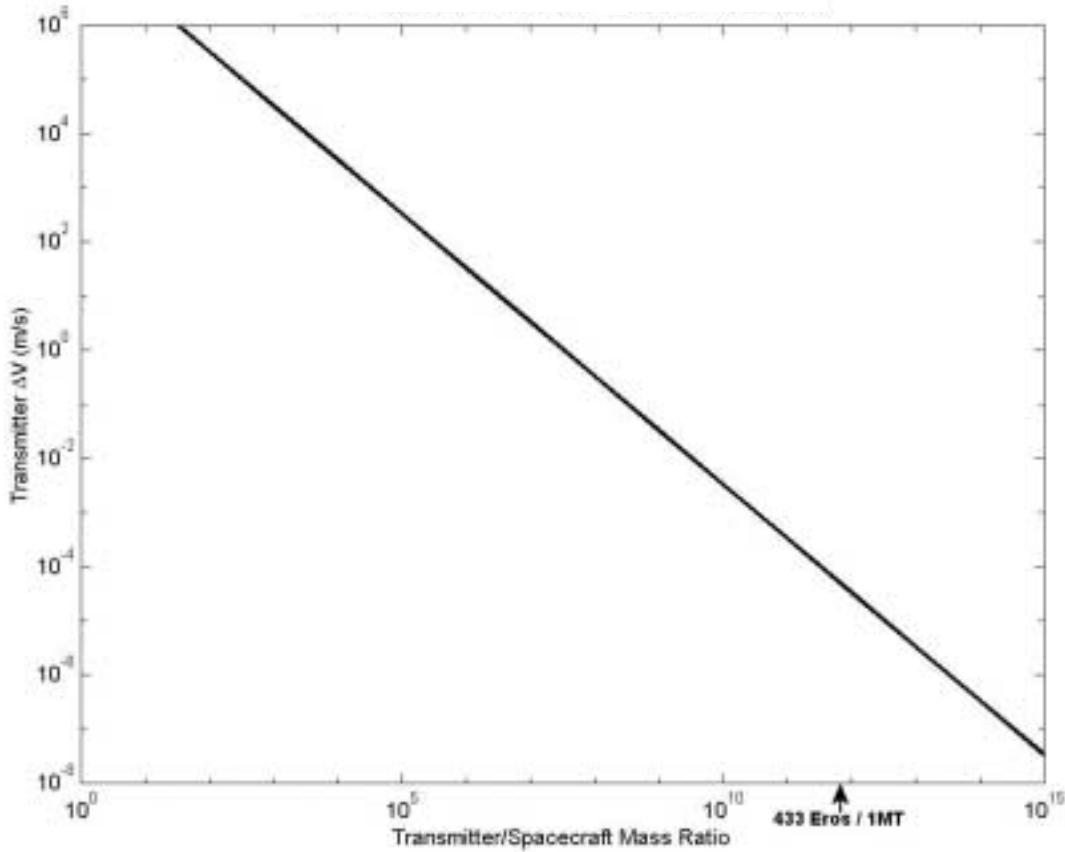

Figure 2: Transmitter ΔV vs. Transmitter/Spacecraft Mass Ratio for a 0.11c spacecraft final velocity

Table 1 gives the acceleration time and distance for the different power consumption lines graphed in Figure 1. Power consumption curves are included in figure 1 for transmitters with sufficient range to reach from Earth to both Mars and Neptune (the outermost major planet[*]). The usefulness of an impulse gravity beam system able to affect spacecraft orbiting neighboring planets could motivate the development of a transmitter with such interplanetary range, even before it was needed for interstellar propulsion.

Judging from the information shown in Figure 1 and Table 1, acceleration on the order of 100 gees over a distance on the order of $4 \times 10^{11}$ meters (roughly the maximum distance from the Earth to Mars) looks like an attractive possible operating regime for launching early interstellar spacecraft missions. As will be discussed later, there are compelling reasons to build an impulse gravity beam transmitter near or on Earth capable of effectively propelling spacecraft at this distance. If the target star system was approximately along the ecliptic plane, then the spacecraft launch could be scheduled so that a second impulse gravity transmitter located near Mars (constructed, perhaps, as part of an Earth-Mars space transportation infrastructure) could assist in accelerating the spacecraft. As Table 1 shows, the acceleration time for a spacecraft being propelled in such a manner is approximately ¼ of a day, suggesting that it might be possible to schedule the spacecraft launch so that it is nighttime in one or more of the regions sponsoring the mission. That way, an impulse gravity generator on or near Earth might be able to receive some or all of its power from "off peak demand" electrical power from the region's electrical power grid. While Figure 1 illustrates that even at high system efficiencies, the world does not generate enough power to launch the example interstellar mission using excess generation capacity, presumably global power production capacity will have increased greatly by the time such a mission is being seriously considered.

---

[*] our sincerest apologies to Pluto

The energy expended in a propulsion system is intended to provide a velocity change to the target spacecraft, however Newton's second law of motion indicates that the beam transmitter will experience a back reaction that will also cause it undergo a velocity change in the direction opposite that of the spacecraft's, as described by the equation:

$$m_t \Delta v_t = m_s \Delta v_s \tag{5}$$

where $m_t$ is the mass of the transmitting object, $\Delta v_t$ is the velocity change of the transmitting object, $m_s$ is the mass of the target spacecraft, $\Delta v_s$ is the velocity change of the target spacecraft, and assuming that the objects are not affected by any other significant forces such as solar wind, maneuvering thrusters, or gravitational attraction to an orbited body. Figure 2 shows the transmitter velocity change that will be experienced by a single transmitter propelling the example 1 MT interstellar spacecraft to a velocity of 0.11c, as a function of transmitter/spacecraft mass ratio. While the transmitter and associated power handling/generation equipment is likely to be much heavier than the launched spacecraft, the back reaction on a space-based transmitter is still significant and needs to be taken into account. For beamed propulsion concepts that operate at low accelerations over long periods of time, the problem of managing this back-reaction may be solved by using solar pressure or unbalanced gravitational forces to offset the reaction forces. Because the impulse gravity generator is likely to operate at higher accelerations, such methods of counterbalancing the reaction force may not be applicable or may require long delays between launches to reposition the transmitter. The back reaction problem might be solved, or at least greatly reduced, by scheduling spacecraft launches in separate directions so that the back reaction of one launch helps to cancel out the effects of the previous launch. If the transmitter can be considered disposable, then it may be unnecessary to even worry about managing the velocity change caused by spacecraft launches, and the transmitter might be simply abandoned after it has been accelerated away from any useful locations.

If the transmitter is in orbit around a celestial body, and the acceleration time is short relative to the orbital period, then the back reaction might be used to stop and then reverse the direction of the transmitter's orbit. In this way, the transmitter can remain in a useful location even with a fairly large change in velocity. A transmitter propelling the sample 1 MT interstellar probe would need to have a mass on the order of 1000 MT to be able to perform such an orbital reversing maneuver while in most Earth orbits.

The velocity imparted on a transmitter due to back reaction forces might be reduced considerably by making the transmitter very massive. This may be necessary simply because of the large power requirements interstellar launches would require. The transmitter could also be made more massive by anchoring it to a large celestial body. To illustrate that even a large asteroid is sufficient to almost eliminate transmitter velocity changes figure 2 indicates the transmitter/spacecraft mass ratio that would be achieved by a transmitter anchored to the asteroid 433 Eros pushing a spacecraft with a mass of 1 MT. Besides asteroids, other likely locations for anchoring such a transmitter might be on Mercury (if the transmitter is solar powered), the Moon, or even the Earth. Because a transmitter located on Earth would have to operate through the atmosphere, whether or not an Earth-based transmitter would be practical would depend heavily on how low atmospheric losses to beam power would be and how much advantage would be achieved by having ready access to Earth's infrastructure. Transmitters located on Earth, or any other rotating bodies, will likely be limited to high power missions with short acceleration times, because of the limited time which the transmitter could keep the spacecraft in its field of view.

Interplanetary

Interplanetary spacecraft require a considerable velocity change to travel from Earth to their destination. This velocity increase is currently provided by conventional rocket boosters at great expense. Using beamed propulsion to provide some or all of the velocity needed for interplanetary spacecraft might allow more difficult or more frequent missions to be accomplished with greater economy. If the Earth-to-orbit impulse gravity beamed propulsion system described in the next section can be constructed then it may be possible to utilize the same system to push the spacecraft past Earth escape velocity and put it on an interplanetary trajectory. Alternatively, a conventional launch system might be used to place the spacecraft in a low-Earth parking orbit, and a space-based impulse gravity beamed propulsion transmitter used to accelerate the spacecraft from the parking orbit onto its interplanetary departure trajectory.

Using an impulse gravity transmitter to accelerate a 1 MT spacecraft from a 230 km altitude parking orbit above Earth by 21.7 km/s would put it on an interplanetary departure trajectory with a hyperbolic excess velocity of 27.4 km/s,[*] and would require approximately $2.4 \times 10^{11}$ J of energy to be imparted to the spacecraft in the form of kinetic energy. This considerable spacecraft velocity would permit a wide range of ambitious interplanetary science missions. The same amount of kinetic energy increase would accelerate a 25 MT spacecraft by 4.3 km/s, providing a hyperbolic excess velocity of 5.1 km/s if started from the same 230 km altitude parking orbit. If the impulse gravity beam can be safely used to propel manned spacecraft, then this 25 MT mission profile would be very useful for propelling manned crew transfer vehicles to Mars. Figure 3 shows the average impulse gravity transmitter power requirements for both of these interplanetary mission types as a function of system efficiency for several different acceleration times.

---

[*] This corresponds to a C3 of 750km$^2$/s$^2$.

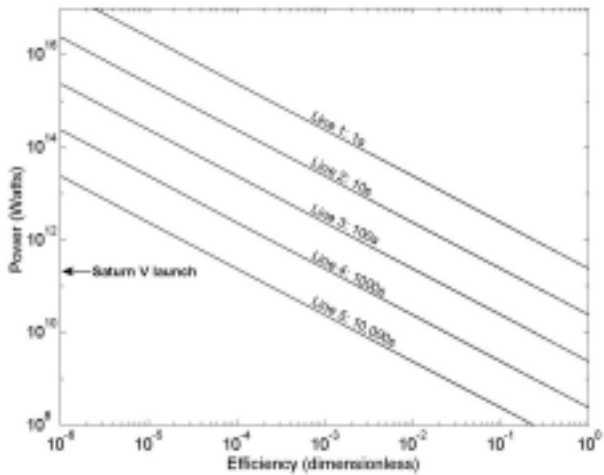

Figure 3: Power Required vs. System Efficiency and Acceleration Time for a 1MT spacecraft accelerated by 21.7 km/s or a 25MT spacecraft accelerated by 4.3 km/s

|      | Acceleration Time |        | Average Acceleration (gees) |              |
| ---- | ----------------- | ------ | --------------------------- | ------------ |
| Line | (s)               | (hrs)  | by 4.3km/s                  | by 21.7km/s  |
| 1    | 1                 | 0.0003 | 438                         | 2217         |
| 2    | 10                | 0.0028 | 43.8                        | 221.7        |
| 3    | 100               | 0.028  | 4.38                        | 22.17        |
| 4    | 1000              | 0.28   | 0.438                       | 2.22         |
| 5    | 10000             | 2.8    | 0.044                       | 0.22         |

Table 2: Average Acceleration for Spacecraft Velocity Increases of 4.3km/s and 21.7 km/s

The power generated by a Saturn V first stage is also shown in figure 3 for comparison. Table 2 shows the required average acceleration for the different acceleration times graphed in figure 3. The very short acceleration times for some of the lines in figure 3 suggest that a high power interplanetary impulse gravity beamed propulsion system might be powered by energy accumulated in storage devices, such as flywheels or capacitors, and released quickly. Low power, long acceleration time impulse gravity beam systems might also be used in these sample applications, but would have greater problems in tracking the target spacecraft and maintaining line of sight if the transmitter were on Earth or in Earth orbit.

Figure 4 shows the back reaction force on the transmitter as a function of transmitter mass for the two example interplanetary missions discussed above. For transmitters that are built on the Earth, the Moon, or some other convenient celestial body, the transmitter velocity resulting from the back reaction on the transmitter for these interplanetary missions is negligible.

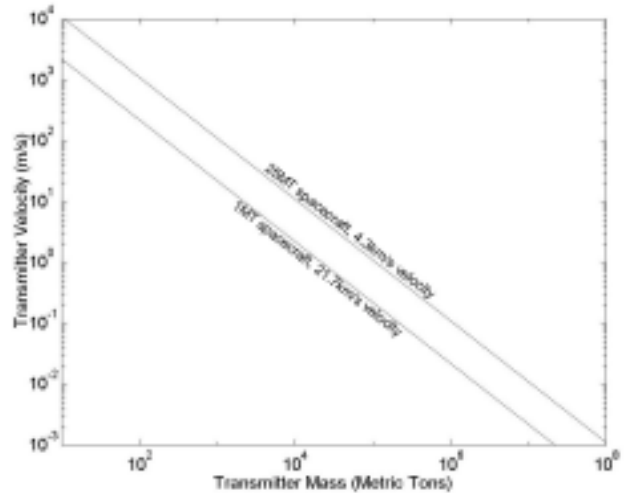

Figure 4: Transmitter Velocity vs. Transmitter Mass for Sample Interplanetary Missions

For an impulse gravity transmitter in Earth orbit, however, the velocity imparted to the transmitter can be quite significant. If the orbiting transmitter platform has a mass of 500 MT, then it would experience a velocity change of 217 m/s and 43 m/s for the 25 MT and 1 MT sample spacecraft missions, respectively. Presumably, any such transmitter would be designed as a reusable piece of space infrastructure and would need to have this velocity change corrected in order to remain in its design orbit. If the transmitter velocity change must be made with conventional means, such as the consumption of rocket fuel brought up from Earth, then it seems unlikely that there would be much advantage to this propulsion system as a means of launching interplanetary spacecraft. In that case, the need for conventional propulsion has not been eliminated, it has only been moved to another portion of the system. If a more economical method can be used to correct the orbital velocity of an impulse gravity transmitter platform, such as the use of an electrodynamic tether system, then an orbiting transmitter may still be a viable method of providing propulsion to interplanetary spacecraft during the Earth departure portion of their mission.

Perhaps the most useful interplanetary application of an impulse gravity beamed propulsion system would come after the interplanetary spacecraft arrived at its destination planet. If a system can be constructed with an effective range long enough to reach from Earth to spacecraft orbiting other planets in the solar system, then beamed propulsion could be used to provide some of the velocity change needed for those spacecraft to make orbital maneuvers at their destination planet. Using impulse gravity beamed propulsion in this manner could reduce the amount of propellant that the spacecraft must carry for its maneuvering thrusters, or allow longer mission times with the same amount of propellant. It would be very useful if the system could help decelerate a spacecraft upon its arrival at a destination planet to permit planetary orbital insertion without expending a large amount of propellant or requiring an aerobraking maneuver. Unfortunately, it may not be possible to accomplish this with an impulse gravity beamed propulsion system, for the same reasons that it may not be possible to use it for decelerating interstellar spacecraft.

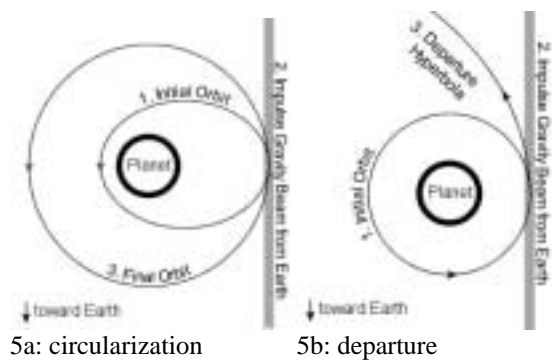

5a: circularization    5b: departure

Figure 5: Maneuvering of Interplanetary Spacecraft Using an Impulse Gravity Beam from Earth

Figure 5a shows, however, how the system might be used to help circularize a spacecraft's orbit after orbital insertion at the destination planet has been accomplished. In addition to helping the spacecraft change its orbit at a destination planet, an impulse gravity beamed propulsion system with interplanetary range would be able to push a spacecraft out of orbit of the destination planet and on to an interplanetary transfer trajectory for a follow-on destination, as shown in figure 5b. Using this method, a spacecraft could conduct a tour of several planets with long loiter times at each destination.

Earth to Orbit
The high cost of launching payloads into orbit with current systems is a considerable impediment to further exploration and exploitation of space resources. Using beamed propulsion to put payloads into Earth orbit has been proposed by numerous previous technical papers,[17,18] but has yet to be accomplished in practice. It is unknown at this time if an impulse gravity based Earth-to-orbit propulsion system can be economically competitive with other launch concepts, but there is no current evidence that it cannot be. If an impulse gravity beamed propulsion transmitter and tracking system is constructed on Earth then it might also be used for more applications than just putting payloads in orbit. An Earth-based impulse gravity transmitter could be used as a component in any of the other three propulsion applications described in this paper, even including interstellar missions. Possible inefficiency due to atmosphere effects, pointing accuracy and field of view limitations resulting from the Earth's rotation, and/or political considerations may hamper the use of ground based impulse gravity beamed propulsion transmitters for any application. On the other hand, low construction and operating costs due to easy access to Earth's infrastructure combined with a stable platform to manage back reaction forces may make ground based transmitters an attractive option even if they suffer some technical limitations relative to space based transmitters. Ironically, this means that some of the power providing propulsion for our most ambitious future space projects may be generated at coal burning power plants operating on pre-World War 2 technologies

One of the problems with using an exclusively ground based impulse gravity beamed propulsion system for launching payloads into orbit is that the payload can only be accelerated directly away from the beam transmitter. While getting both the altitude and velocity needed for orbit with such constraints is difficult, it is not impossible. In his paper "Trajectory Simulation for Laser Launching" J.T. Kare describes how this could be accomplished by launching the payload "on a sub-orbital path and allowing it to coast until it has a downward velocity,"[19] and then applying a second acceleration phase to the payload with either the original transmitter, or a second transmitter further downrange. These trajectories are shown in Figure 6.

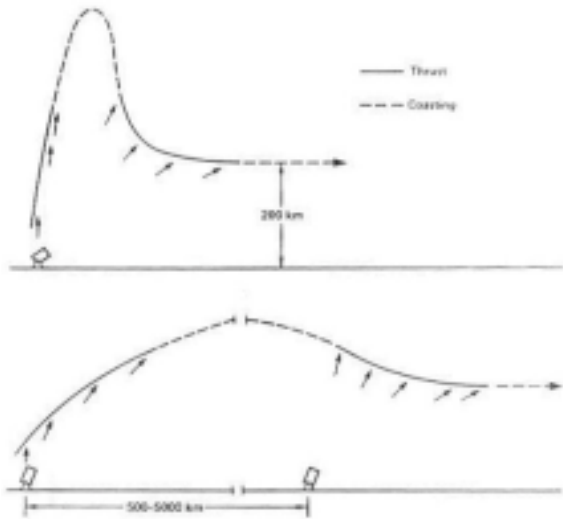

Figure 6: Methods for Launching a Payload Into Orbit With the Thrust Constrained Along the Transmitter-Payload Axis[19]

One of the most obvious differences between the Earth-to-orbit application and previously discussed impulse gravity beamed propulsion concepts is the presence of a large amount of intervening matter between the transmitter and target in the form of the Earth's atmosphere. The exact amount of air between a ground based impulse gravity generator and a target being launched into (or beyond) Earth orbit is dependent on the diameter of the impulse gravity beam, altitude of the target, the altitude of the transmitter, angle of the beam's travel through the atmosphere, and even local weather conditions. Figure 7 shows the approximate mass of air that would exist in a beam fired straight up through the atmosphere from sea level. This figure gives some sense of scale for how much intervening air may exist between the transmitter and target spacecraft in an Earth-to-Orbit beamed propulsion system.

To have a beam diameter sufficient to encompass the entire target spacecraft throughout the launch trajectory using a constant diameter beam would likely require at least a 100 inch (2.54 m) beam diameter for typical medium sized (<5 MT) payloads and a 250 inch (6.35 m) beam diameter for typical large ($\cong$30 MT) payloads. It can be seen from figure 7 that with these beam diameter and payload assumptions, the mass of intervening air between the transmitter and target in the later portions of an Earth-to-orbit launch can be an order of magnitude greater than the mass of the target payload. If the intervening air absorbs the energy of the impulse gravity beam as effectively as the matter comprising the target payload, then the mass of the air in the beam path will be the dominant factor in beam power requirements for such a system. Even if, as suggested in the "Impulse Gravity Generator Theory" subsection, the intervening air does not absorb the impulse gravity beam as effectively as solid objects it could still be an important factor in beam power requirements because the mass of intervening air is so large relative to the payload masses.

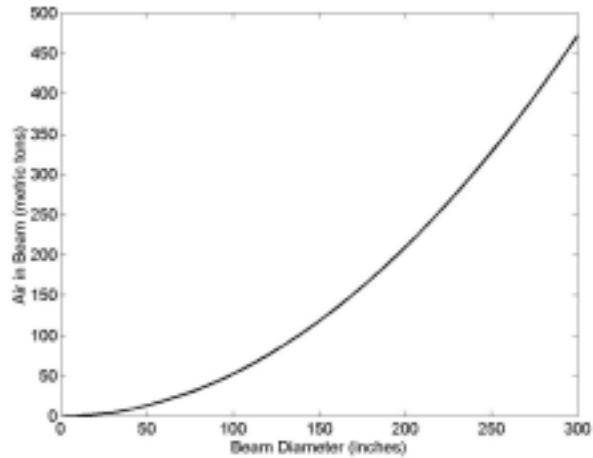

Figure 7: Air in a Cylindrical Vertical Column Extending from Sea Level to Space vs. Cylinder Diameter

There are several possible ways that the amount of air in the beam path between the transmitter and target might be reduced. One way to accomplish this would be to reduce the beam diameter. Beam diameter could be reduced while still remaining large enough to contain the entire target payload by increasing the pointing accuracy of the beam and/or decreasing the diameter of the payload as seen by beam. Target diameter could be reduced either by increasing the packing density of the payload, or by having a target with a large fineness ratio and keeping the long axis of it pointed toward the transmitter. In the later portions of the launch trajectory where the target payload is at a relatively large distance from the transmitter, a small diameter transmitter combined with the proper beam divergence could allow the beam diameter to remain small in the lower, dense sections of the atmosphere while still being large enough at the higher altitudes to allow the entire target to remain within the beam. A beam fired straight upward with an initial diameter of 0.01 m and a divergence of 2 arc-seconds would only have a beam diameter of 0.21 m by the time it exits the troposphere (11,000 m altitude) but would have spread to a diameter of 6.7 m at an altitude of 230 km.

Using a beam that intersects only a portion of the target payload could also reduce the beam diameter. A beam that intersects only part of the target would cause it to experience additional internal stresses arising from the differences between the acceleration provided by the beam on matter in the beam path and the normal gravitational forces applied to matter outside the beam. This might not be a problem when launching bulk material but it could damage sensitive spacecraft components. Directing a small diameter beam to one or more inert target masses that are then connected to the payload by a load carrying structure might allow delicate payloads to be propelled by a small diameter beam without suffering damage from internal stresses.

Another method of reducing the amount of air between the transmitter and target would be to put the transmitter at the top of a mountain, as is currently done with many astronomical observatories, to get it above as much of the atmosphere as possible even before the launch occurs.

The specific energy required to send a payload to a 230 km altitude orbit using a "perfect" 100% efficient launch system is approximately $3.2 \times 10^6$ J/kg. The specific energy required for a real launch system to place a payload in orbit can described by the equation:

$$\varepsilon_{real} = \eta_{launch} \varepsilon_{theoretical}$$

(6)

where $\varepsilon_{real}$ is the specific energy consumed by the real launch system, $\eta_{launch}$ is the overall efficiency of the launch system, and $\varepsilon_{thoretical}$ is the specific energy increase in the payload or the amount of energy required per unit mass of payload for a "perfect" 100% efficient launch system. The term $\eta_{launch}$ in equation 6 will be smaller than the term $\eta_{system}$ used in equation 3, because in addition to including the inefficiency in the impulse gravity beamed propulsion system, $\eta_{launch}$ also considers inefficiencies arising from other parts of the launch process, such as air friction, trajectory related losses, and intervening matter in the beam path. Since the form of energy used to power an impulse gravity beam based Earth-to-orbit launch system would come from electricity, the energy cost per unit mass of payload can be calculated by multiplying $\varepsilon_{real}$ by the price of electricity. Figure 8 shows both the energy required and energy cost per kilogram for an impulse gravity beam based Earth-to-orbit launch system as a function of overall launch system efficiency assuming an electricity cost of $1.4 \times 10^{-8}$ per Joule (5¢/kw*h).

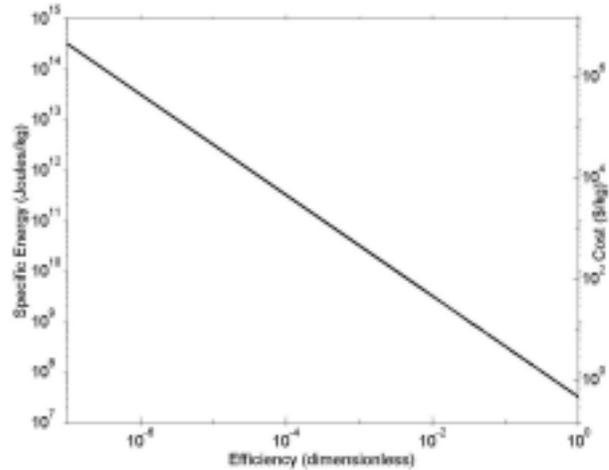

Figure 8: Energy and Energy Related Cost to Launch 1kg into a 230km Orbit vs. Launch System Efficiency

It can be seen from Figure 8 that to achieve an energy cost of $220/kg ($100/lb) with the assumptions given above would require a launch system efficiency of approximately 0.002 (0.2%). With the current information about impulse gravity beam behavior, it is impossible to predict how soon such efficiencies will be attainable.

Low energy cost for a launch system does not necessarily mean a low overall launch cost; after all, fuel cost is only a small portion of the Space Shuttle's launch cost. To be able to launch payloads cheaply to low Earth orbit, a system also needs features such as a high degree of reusability, low capital cost, and low labor and overhead costs. While the economics of an impulse gravity beam based Earth-to-orbit launch system cannot by quantified at this time, it would appear to have many qualities indicative of low cost operations. With the possible exception of an aerodynamic shell to cover the payload, nothing in the launch system would be expendable. Critical propulsion components will remain on the ground, allowing them to be made to less exacting weight and space requirements, and also permitting easy inspection and repair. Because the payload does not need to be attached to the top of large containers of energetic fuels and oxidizers, as with a conventional rocket based launch system, it may be possible to recover the payload intact in the event of a launch system malfunction, reducing launch related insurance costs for expensive payloads and decreasing the political resistance to launching hazardous materials. Some of these benefits are common to other beamed propulsion Earth-to-orbit systems, not just ones based on impulse gravity beams.

All of the impulse gravity beamed propulsion applications described in this paper lend themselves to being combined with other propulsion systems and the Earth-to-orbit application is no exception. Even if a purely impulse gravity based Earth-to-orbit system is not competitive, it may be practical as part of a hybrid launch system using a mixture of beamed propulsion and more conventional propulsion. For example, space based or Earth based impulse gravity transmitters might act as an upper stage that could provide propulsion and/or circularize the orbit of gun launched payloads, eliminating the need for the gun launched shells to have internal propulsion and guidance systems. A large short range impulse gravity transmitter built into a launch pad might be used to assist the first stage propulsion of large launch vehicles with low thrust to weight ratios, reducing the amount of time it takes such vehicles to clear the launch tower and/or reducing the need for strap-on boosters. A launch pad based impulse gravity system might also be an enabling technology for fully reusable single-stage-to-orbit launch vehicles by helping provide the propulsion needed for an otherwise marginally powered design to safely reach orbit, lowering the amount of technical innovation that must be incorporated into the launch vehicle itself.

Orbital Maneuver near Earth

A beamed propulsion system based on an impulse gravity beam would be especially useful for orbital maneuver and orbit raising of Earth orbiting satellites because it would require no specialized propulsion components on the target spacecraft. No additional mass would need to be added to new spacecraft designs to take advantage of such a system; it could be used on existing spacecraft without requiring them to undergo any modification and on non-cooperative targets like space debris or 'dead' satellites. Unlike the interplanetary orbital maneuver application described previously, a combination of ground and space based transmitters could allow target spacecraft to be propelled in any direction needed. This impulse gravity transmitter network could completely eliminate the need for the target spacecraft to use its own onboard propulsion system. If high enough beam accuracy and control could be achieved, it may even be possible to use the system to apply a torque to the spacecraft with a pair of impulse gravity beams directed at different portions of the target spacecraft from opposing directions in order to allow the target spacecraft to desaturate its reaction wheels without using onboard propulsion or pointing systems.

In order to construct a system of impulse gravity based beamed propulsion transmitters that could provide orbital maneuver propulsion to spacecraft in Earth orbit, it would be necessary to place some of the transmitters in a constellation of spacecraft with orbital altitudes higher than the expected target spacecraft. Figure 9 shows how an arrangement of three ground based and one space based transmitter could be used to provide propulsion in any direction to a target spacecraft between them. If an attractive impulse gravity beam transmitter could be developed, then the space-based transmitters could be eliminated without reducing the omnidirectional capability of the network, however there is no current evidence that an attractive impulse gravity beam can be constructed.

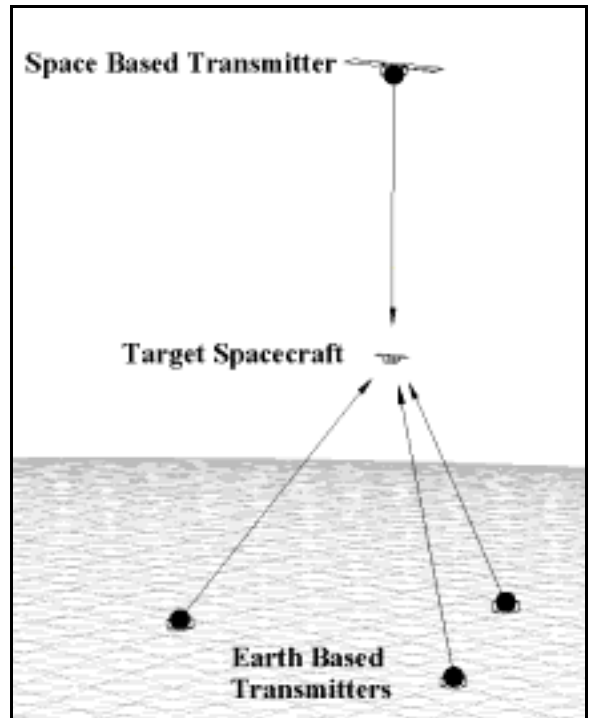

Figure 9: Four Impulse Gravity Beamed Propulsion Transmitters Able to Propel Target Spacecraft in Any Direction (not to scale)

Back reaction on the transmitter from the impulse gravity beam will require the space-based transmitters to use conventional spacecraft propulsion to balance the back reaction and maintain their orbits when the transmitter is in use. An impulse gravity beamed propulsion network for satellite orbital maneuvering that contains space-based transmitters does not, therefore, completely eliminate the need for conventional spacecraft propulsion, it merely transfers that burden from the target spacecraft to the transmitter spacecraft.

Despite the need for the transmitter spacecraft to use more conventional propulsion to balance back reaction forces, there may still be a net gain achieved by building and operating an impulse gravity beamed propulsion network to provide propulsion for spacecraft and other objects near Earth. If the system efficiency is sufficiently high and the geometry of the target satellite and the transmitters can be arranged so that the ground based transmitters provide most of the propulsion needed, then the amount of fuel it consumes could be less than the fuel that would have been consumed by the target spacecraft. Because they could be designed as specialized "tugboat" spacecraft, the space-based transmitters could be designed to incorporate the most efficient maneuvering thrusters available, allowing them to minimize fuel consumption. Depending on system efficiency, it might be cheaper to install high efficiency thrusters on just a few purpose built transmitter spacecraft that could provide propulsion to many others than it would be to install equivalently efficient thrusters on many different regular spacecraft. Because they would be optimized for propulsion duties the space-based transmitters might also be designed to be refueled on-orbit, or to simply be built with very large fuel tanks and then disposed of when empty. The transmitter spacecraft might even make use of other propellantless propulsion methods, such as tethers.

There are some types of satellites, such as Earth observation platforms, that require frequent repositioning and whose useful life is limited by the amount of onboard fuel. If these satellites could be repositioned with a beamed propulsion network instead of consuming onboard fuel then their useful life might be increased considerably, possibly resulting in an overall cost savings if the value of the target satellites is high relative to the cost of the beamed propulsion network. Even if there is no overall cost savings due to reduced fuel use the system may still be beneficial for use with satellites mounting delicate components that could be damaged by exhaust gasses from onboard thrusters.

If the availability of a beamed propulsion network could reduce the cost of conducting orbital maneuvers then it might allow new strategies for satellite use. "Roving satellites" that undergo considerable orbital changes to provide temporary increases in regional capacities might become common, such as weather satellites that could be moved to better track emerging storms as they develop or communication satellites that could be moved to provide temporary increases in bandwidth during high profile events. It might also allow future satellite constellations to fill holes caused by satellite outages quickly and cheaply by having a smaller number of spare satellites stored on orbit that could undergo very large orbital changes with the assistance of the beamed propulsion network to reach any point in the constellation they are needed. A space-based transmitter would also be useful in deorbiting old satellites and space debris.

Even if the space-based transmitter portion of the network was not practical, a partial network composed only of ground-based transmitters could still assist spacecraft in conducting orbital maneuvers and reduce the amount of fuel consumed by the target spacecraft. A ground-based network would also be useful in helping satellites to avoid collisions with space debris by moving the debris to a less hazardous orbit. The most potentially profitable use that a ground-based impulse gravity beamed propulsion network could be put to would likely be to boost satellites from low-Earth-orbits to geostationary orbits. The exact size of the market for low-Earth-orbit to geostationary orbit orbital transfer is unknown, but figure 10 shows the potential gross revenue for likely ranges of price and launch volume over the next couple of decades. Assuming that a fee of approximately $1000/kg ($454/lb) of satellite mass (significantly less than current low-Earth-orbit to geostationary orbit costs[20]) could be charged and annual launch rates of geostationary satellites were on the order of 100,000 kg/year (consistent with currently projected rates for the next decade[21]), then approximately $100,000,000 per year of gross revenue could be expected from this service.

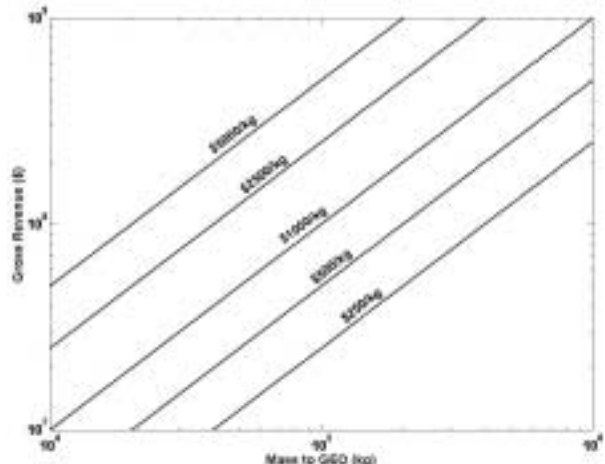

Figure 10: Potential Gross Revenue from Boosting Satellites from LEO to GEO vs. Mass Boosted and Price per kg

Capital and operating costs for the proposed network of ground-based transmitters is unknown. Considering the large potential value of the orbit boosting, collision prevention, satellite disposal, and orbital maneuver assistance services that such a network could provide it seems likely, however, that if the technology can be developed then the construction of a ground-based impulse gravity beamed propulsion network for near Earth use could be economically justifiable. Addition of space-based transmitters would expand the capabilities of the network even further. Success of impulse gravity based beamed propulsion for near Earth orbital maneuver could pave the way for its use in the more ambitious applications previously discussed.

Topics Needing Further Study

The previous examples demonstrate that an impulse gravity generator based beamed propulsion system could be an extremely useful tool in the exploration and exploitation of outer space. Unfortunately, our understanding of impulse gravity beams is not sufficiently developed at this time to allow us to predict the exact characteristics of a mature propulsion system based on them. In order to determine whether or not the vast potential of impulse gravity based beamed propulsion systems can actually be realized, more empirical data must be collected experimentally, more complete theoretical models must be developed, and related technologies must be refined. Some of the areas of research that need further study are:

- More accurate measurements of impulse gravity beam divergence over long distances
- Determining if and/or how beam divergence can be controlled
- Determining if and/or how steady state impulse gravity beams can be generated
- Measuring beam absorption in and effect on air and other fluids
- Determining theoretical limits of and practical ways to improve system efficiency
- Developing ways to decrease cost and improve operation of impulse gravity generators
- Experimental determination of the effect of beam exposure to delicate cargo, including biological specimens
- Understanding system behavior under beam power limited conditions due to massive and/or fast target masses
- Determining if beam cross sectional area occupied by the target effects energy absorption from the beam (i.e. Can the beam be characterized as having an energy per unit of beam cross sectional area?)
- Determining theoretical limits of and practical ways to improve beam pointing accuracy and/or decrease jitter.
- Determining if and/or how an impulse gravity beam could be focused, reflected, or otherwise manipulated
- Developing quantitative empirical and theoretical models that would allow conceptual design, sizing, and costing of future impulse gravity based beamed propulsion systems.

Conclusions

Impulse gravity generators having the characteristics described by Podkletnov and Modanese[6] have considerable potential for use in a number of beamed propulsion applications. These applications range from near term commercial uses such as boosting satellites from low-Earth-orbit to geostationary orbit, to far term exploratory missions to neighboring stars. Likely advantages of an impulse gravity based beamed propulsion system include:
- Lack of a need for specialized propulsion components onboard the spacecraft (or other object) being propelled.
- Relatively high efficiency for a beamed propulsion system at any target velocity
- Low operating cost
- Ability to propel spacecraft at high acceleration while producing little or no internal stresses

Disadvantages to impulse gravity beamed propulsion may include:
- Poor accuracy, possibly resulting in short operational range
- Not well suited to providing remote spacecraft deceleration
- Propulsion can only be provided in one direction, constrained along the transmitter-target axis

Unfortunately, the technology and theoretical models dealing with impulse gravity beams are still in the early stages of development. A great deal of further study is required to determine if and how the promise of impulse gravity beamed propulsion can be realized.

APPENDIX A: amplitude of the anomalous gravitational fluctuations in superconductors
(see "Impulse Gravity Generator Theory" subsection)

The amplitude of the anomalous fluctuations [VacNew] is proportional to $|L - \Lambda/8\pi G|$, where $L$ is the lagrangian of the Cooper pairs wave function (Ginzburg-Landau lagrangian) and $\Lambda$ is the background value of the vacuum energy density at the scale of interest (~ $10^{-8}$ cm). At cosmological scale $\Lambda$ is known to be negative and of the order of $10^{-50}$ $cm^{-2}$, so that $|\Lambda/8\pi G| \sim 10^{-1}$ $J/m^3$ in SI units. However, $\Lambda$ is probably scale-dependent, and increases at small distances (see [VacNew]).

We have shown that $L$ can be expressed in a simple form, in which the magnetic field does not appear explicitly, as a function of the Cooper pairs density only:

$$L = -\frac{1}{2m}\left[\hbar^2(\nabla\rho)^2 + \hbar^2\rho\nabla^2\rho - m\beta\rho^4\right]$$

where $m$ is the pair mass, the density of Cooper pairs is equal to $\rho^2(x)$ and $\beta$ is linked to the value of the Ginzburg-Landau parameter $\kappa=\lambda/\xi$ by the relation $\kappa^2 = m^2\beta/(2\mu_0\hbar^2 e^2)$. It is straightforward to check that the sign of $L$ is positive for two types of configurations:

(1) For constant solutions of the Ginzburg-Landau equation in the absence of external field, which implies $\rho^2(x) = n_p$, being $n_p$ the average pairs density. The corresponding constant lagrangian density is

$$L_1 = \frac{1}{2}\beta n_p^2.$$

(2) For regions of the superconductor where $\rho\nabla^2\rho$ is negative and greater, in absolute value, than $(\nabla\rho)^2$. It is straightforward to check that these are regions located around local density maximums, or more generally lines and surfaces where the first partial derivatives of $\rho$ are zero and the second derivatives are negative or null. The lagrangian density at a maximum is $L_2 \sim \frac{\hbar^2}{2m}\rho|\rho''|$. If the maximum is sharp, $L_2$ can be much larger than $L_1$. Configurations of this kind are characteristic of solutions of the Ginzburg-Landau equation with strong magnetic flux penetration [Tilley].

In all other configurations, $L$ is negative. Some numerical estimates are given in Table 1, where the gradients are taken to be of the order of $\rho/\xi$. At local minima or in regions with strong gradients, we can suppose that $|L|$ is of the same magnitude order as $L_2$.

|  | Pb | YBCO |
|---|---|---|
| $\lambda$ (m) | $3.9 \cdot 10^{-8}$ | $1.4 \cdot 10^{-7}$ |
| $\xi$ (m) | $8.2 \cdot 10^{-8}$ | $1.6 \cdot 10^{-9}$ (a-b direction) |
|  |  | $2.4 \cdot 10^{-10}$ (c direction) |
| $n_p$ ($m^{-3}$) | $9.3 \cdot 10^{27}$ | $3.5 \cdot 10^{27}$ |
| $L_1$ (J $m^{-3}$) | $10^4$ | $10^6$ |
| $L_2$ (J $m^{-3}$) | $10^4$ | $10^6$ (a-b direction) |
|  |  | $10^8$ (c direction) |

*Magnitude orders of the lagrangian densities $L_1$ and $L_2$ for a type I superconductor (Pb) and for a type II superconductor (YBCO), computed according to the equation above. Also listed are the values of the London length $\lambda$ and the coherence length $\xi$ at T=0 [Waldram] and the average pairs density $n_p$, computed from $\lambda$ by the relation $n_p = m_e/(2\mu_0 e^2\lambda^2)$. For YBCO, the values of $\xi$ along the a-b direction and the c direction are given separately, and so the corresponding values of $L_2$; $n_p$ is the density in the a-b planes, computed with the effective mass $m^*=4.5 m_e$.*

[VacNew] G. Modanese, "Local contribution of a quantum condensate to the vacuum energy density", LANL Physics Preprint Server, preprint gr-qc/0107073.

[Tilley] D.R. Tilley and J. Tilley, "Superfluidity and superconductivity", IoP, Bristol, 1990.

[Waldram] J. Waldram, "Superconductivity of metals and cuprates", IoP, London, 1996.

[VacNew] G. Modanese, "Local contribution of a quantum condensate to the vacuum energy density", LANL Physics Preprint Server, preprint gr-qc/0107073.

[Tilley] D.R. Tilley and J. Tilley, "Superfluidity and superconductivity", IoP, Bristol, 1990.

[Waldram] J. Waldram, "Superconductivity of metals and cuprates", IoP, London, 1996.